\documentclass[pra,twocolumn,floatfix,showpacs,superscriptaddress]{revtex4}

\usepackage{amsmath}
\usepackage{amsfonts}
\usepackage{graphicx}

\DeclareMathSymbol{\N}{\mathalpha}{AMSb}{'116}

\DeclareMathSymbol{\R}{\mathalpha}{AMSb}{'122}

\begin{document}

\title{Non-equilibrium entangled steady state of two independent 
two-level systems}
\author{S. Camalet}
\affiliation{Laboratoire de Physique Th\'eorique de 
la Mati\`ere Condens\'ee, 
UMR 7600, Universit\'e Pierre et Marie Curie, 
Jussieu, Paris-75005, France}
\date{Received: date / Revised version: date }
\begin{abstract}
We determine and study the steady state of two independent 
two-level systems weakly coupled to a stationary non-equilibrium 
environment. Whereas this bipartite state is necessarily 
uncorrelated if the splitting energies of the two-level systems are 
different from each other, it can be entangled if they are equal. 
For identical two-level systems interacting with two bosonic heat 
baths at different temperatures, we discuss the influence of 
the baths temperatures and coupling parameters 
on their entanglement. Geometric properties, such as the baths 
dimensionalities and the distance between the two-level systems, 
are relevant. A regime is found where the steady state is 
a statistical mixture of the product ground state and of 
the entangled singlet state with respective weights 2/3 and 1/3.
\end{abstract} 


\maketitle

\section{Introduction}

For a quantum system, the influence of the surroundings plays a role 
at a fundamental level. When the environment is taken into 
consideration, the system dynamics can no longer be described in 
terms of pure quantum states and unitary evolution. An open quantum 
system is generally in a statistical mixture of pure states. This has 
an important consequence for multipartite systems. As is well known, 
correlations between quantum systems cannot be completely 
understood in classical terms \cite{W}. There exist states which are 
not classically correlated and lead to correlations with no classical 
counterpart, as clearly shown by violations of Bell inequalities for 
instance \cite{B}. They are said to be entangled. Whereas 
almost all pure states are entangled, this is not the case for mixed 
states. In the space of mixed states, the set of non-entangled, or 
separable, states has a finite volume \cite{ZHSL}. An interesting 
consequence of the geometrical properties of this set is that 
the state of a multipartite open system can be entangled for finite 
periods of time, in the course of its evolution, and separable at 
infinite time or vice versa \cite{TC}. 

The most common environment is a heat reservoir. If the considered 
system is weakly coupled to an infinite number of degrees of freedom, 
initially in thermal equilibrium, it relaxes, in general, to a thermal state 
with the temperature of its surroundings. In such an environment, 
it is clear that, in the absence of direct interactions between 
the subsystems of a multipartite system, these subsystems are 
uncorrelated at long times. In other words, any initial correlation, 
quantum or classical, between independent subsystems is 
generically destroyed by a thermal bath. Moreover, for 
the geometric reasons mentioned above, quantum 
disentanglement can occur in a finite time \cite{YE,JJ}. 
Furthermore, the disentangling influence of the environment also 
exists when no energy is exchanged between the system and 
its surroundings, whereas, in this particular case, classical 
correlations can persist \cite{DH,YEPRB}. 

However, when independent systems interact with a common 
environment, the indirect interaction between them, mediated by 
this environment, may have a positive impact on their entanglement. 
Recent results evidence the existence of this influence. It has been 
shown that a transient entanglement, between initially uncorrelated 
systems, can be induced by a thermal bath, for both non-dissipative 
\cite{Br} and dissipative \cite{J,BFP,MCNBF} couplings. It has also 
been obtained that, in the limit of infinitely close non-interacting 
systems, some special entangled states are not affected by 
the environment \cite{ZR}. In this limiting case, the considered 
multipartite open system has not a unique steady state, which is 
exceptional, and hence the entanglement evolution depends on 
the system's initial state.  

In the above cited dynamical studies, the environment is in thermal 
equilibrium and thus a relaxation dynamics towards a unique 
steady state necessarily means decay of correlations, both quantum 
and classical, between non-interacting systems. This may not be 
the case for a non-equilibrium surroundings. Stationary entanglement 
has been found in the presence of particle \cite{HDB,CPA} or energy 
flow \cite{PH,KAM}. However, in these studies, entanglement occurs 
between systems that interact with each other directly, via a two-level 
system, or via strong coupling to a heat bath, and this interaction 
plays an essential part in the development of entanglement. Such 
a strong interaction has been shown to be unnecessary for 
a different kind of non-equilibrium environment \cite{CKS}. 
In the presence of a classical oscillating field, the steady state of 
two two-level atoms, interacting with each other only via weak 
coupling to the electromagnetic vacuum, can be entangled.

\begin{figure}
\centering \includegraphics[width=0.45\textwidth]{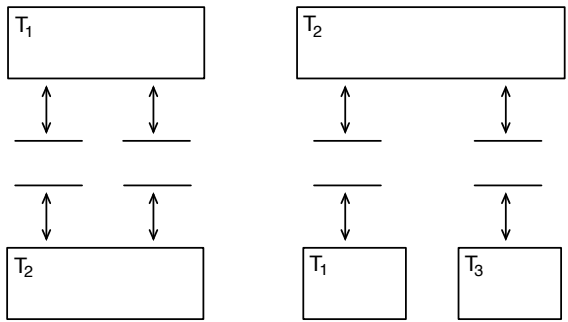}
\caption{\label{fig:test} Schematic representation of non-equilibrium 
environments of two independent TLS. The depicted environments 
consist of several heat baths at different temperatures.}
\end{figure}

In this paper, we consider two independent two-level systems (TLS) 
coupled to a steady non-equilibrium environment. Examples of 
such surroundings are illustrated in Fig.\ref{fig:test}. They consist of 
several heat baths at different temperatures. These are not the only 
possible examples and the two following sections are relevant to 
other environments. In section \ref{sec:M}, we present the model 
used to describe two non-interacting TLS in a stationary environment. 
In section \ref{sec:Ssd}, we first study the steady state of a general 
system weakly coupled to its surroundings, and we then apply 
our approach to the case of a system consisting of two 
independent TLS. The system steady state is obtained, in the weak 
coupling limit, by solving perturbatively an eigenvalue problem, 
which is derived from the system dynamics for arbitrary coupling 
strength. As far as one is interested only in the stationary state, 
no other approximation, such as a Markovian assumption, 
or elaborate method, such as a projection superoperator technique, 
are needed \cite{QDS,CDG}. In section \ref{sec:Bhb}, we focus on 
the special case of an environment that consists of bosonic 
heat baths at different temperatures. It is shown that two baths 
are enough to induce stationnary entanglement of two identical TLS. 
The influence of the two baths temperatures and of the coupling 
parameters is discussed in some detail. Finally, we summarize 
our results in the last section.
         
\section{Model}\label{sec:M}

The total Hamiltonian of two independent TLS and their environment 
${\cal E}$ can be written as
\begin{equation}
H = \sum_{i=1}^2 \left[ -\frac{\Delta_i}{2}  \sigma_z^{(i)} 
+ v_i \sigma_z^{(i)} + w_i \sigma_+^{(i)} 
+ w^{\dag}_i \sigma_-^{(i)} \right] +H_{\cal E}  \label{H}
\end{equation}
where $\Delta_i$ are the level splittings of the TLS, 
$v_i=v_i^{\dag}$ and $w_i$ are operators of ${\cal E}$ and 
$H_{\cal E}$ is the self-Hamiltonian of ${\cal E}$. The Pauli 
operator $\sigma_z^{(i)}$ has eigenvalues $\pm 1$ and 
the corresponding eigenstates are denoted by $| \pm \rangle_i$. 
The operators $\sigma_{\pm}^{(i)}$ then read 
$\sigma_+^{(i)} = [\sigma_-^{(i)}]^{\dag}=| + \rangle_i {_i}\langle - |$.  
We introduce, for further use, the following notations : 
\begin{eqnarray}
| 1 \rangle =| + \rangle_1 | + \rangle_2 &,& 
| 2 \rangle =| + \rangle_1 | - \rangle_2 , \nonumber \\
| 3 \rangle =| - \rangle_1 | + \rangle_2 &,& 
| 4 \rangle =| - \rangle_1 | - \rangle_2. \label{def}
\end{eqnarray}
Two TLS interacting with their environment but not directly with each 
other can always be described by a Hamiltonian of the form \eqref{H}. 
The system ${\cal E}$ is assumed to consist of an infinite number of 
degrees of freedom and to lead to a decohering and dissipative 
reduced dynamics of the TLS. 

As the initial state of the complete system, we consider 
\begin{equation}
\Omega = \sum_{k,l} r_{kl} | k \rangle \langle l | 
\otimes \rho_{\cal E}  \label{Omega}
\end{equation}
where $\rho_{\cal E}$ commutes with $H_{\cal E}$. 
The two-TLS system and ${\cal E}$ are initially uncorrelated. 
As we will see below, the condition $[\rho_{\cal E},H_{\cal E}]=0$ 
implies the stationarity of relevant correlation functions of ${\cal E}$. 
Typical environments we are interested in are made up of several 
heat baths at different temperatures $T_n$, as sketched 
in Fig.\ref{fig:test}. In this case, the environment Hamiltonian 
and initial state read, respectively, as 
$H_{\cal E}=\sum_n H_{{\cal E}n}$ where $n$ runs over 
the heat reservoirs and $[H_{{\cal E}n},H_{{\cal E}n'}]=0$, 
and $\rho_{\cal E}\propto \prod_n \exp(-H_{{\cal E}n}/T_n)$, 
and commute with each other. Throughout this paper, 
we use units in which $\hbar=k_B=1$.

\section{Non-equilibrium steady state}\label{sec:Ssd}

In this section, we first derive a matrix equation for the steady state 
of a generic open system ${\cal S}$ initially uncorrelated with 
its environment ${\cal E}$. More explicit equations are then 
obtained for a steady environment and in the limit of weak coupling 
between ${\cal S}$ and ${\cal E}$. This weak coupling approach is 
applied to the two TLS described by the Hamiltonian \eqref{H}. 
In this case, the steady state equation can be solved. The result 
is radically different for $\Delta_1 \ne \Delta_2$ and 
$\Delta_1 = \Delta_2$.   

\subsection{General case}

In general, under the influence of its environment ${\cal E}$, a system 
${\cal S}$ relaxes to a steady state determined by its self-Hamiltonian 
and by its interaction with ${\cal E}$. If ${\cal E}$ is in thermal 
equilibrium and ${\cal S}$ interacts with it weakly, this state does not 
depend on any detail of the intrinsic dynamics of ${\cal E}$ or of 
the coupling between ${\cal S}$ and ${\cal E}$. But, as we will see, 
this is a very particular case. To determine the steady state of 
${\cal S}$, we first write its reduced density matrix, at positive times 
$t$, as 
\begin{equation}
\rho(t) = \frac{i}{2\pi} \int_{\R+i\eta} dz  e^{-izt}  
\mathrm{Tr}_{\cal E} \left[  \left( z-{\cal L} \right)^{-1} \Omega \right]  
\label{rhot} 
\end{equation}  
where $\mathrm{Tr}_{\cal E}$ denotes the partial trace over 
${\cal E}$, $\eta$ is a positive real number, and $\Omega$ is 
the initial state of the total system ${\cal S}+{\cal E}$. 
The Liouvillian ${\cal L}$ is defined by 
${\cal L} \ldots =[H,\ldots]$ where $H$ is the Hamiltonian 
of ${\cal S}+{\cal E}$. This Hamiltonian can be decomposed as
$H= H_{\cal S} + H_{int} + H_{\cal E}$ where $H_{\cal S}$ and 
$H_{\cal E}$ are the self-Hamiltonians of ${\cal S}$ and ${\cal E}$, 
respectively, and $H_{int}$ accounts for the interaction between 
${\cal S}$ and ${\cal E}$. The condition 
$\mathrm{Tr}_{\cal E} (\rho_{\cal E} H_{int})=0$ can be assumed 
without loss of generality. It can always be satisfied by appropriately 
redefining $H_{\cal S}$ and $H_{int}$. The eigenstates and 
eigenenergies of $H_{\cal S}$ will be denoted by $| k \rangle$ 
and $\epsilon_k$ in the following. 

\subsubsection{Steady state equation}

For an initial state $\Omega$ of the form \eqref{Omega}, 
the matrix elements 
${\tilde r}_{kl}(z) = \langle k | \mathrm{Tr}_{\cal E} 
[  ( z-{\cal L} )^{-1} \Omega ]  | l \rangle$ 
 of the Laplace transform of $\rho$, are given by
\begin{equation}
{\tilde r}_{kl}(z) = \sum_{k',l'} \Gamma_{kl,k'l'} (z) r_{k'l'} 
\label{Meq}
\end{equation}  
where the functions
\begin{equation}
\Gamma_{kl,k'l'} (z) = \langle k |  
\mathrm{Tr}_{\cal E} \left[  \left( z-{\cal L} \right)^{-1} 
| k' \rangle \langle l' | \otimes  \rho_{\cal E} \right] | l \rangle 
\label{Gamma}
\end{equation}  
depend only on the environment part of the initial state \eqref{Omega}.
Equation \eqref{Meq} can be read as a matrix relation between two 
column vectors ${\bf r}$ and ${\tilde {\bf r}}(z)$ with elements $r_{kl}$ 
and ${\tilde r}_{kl}(z)$, respectively, and a square matrix 
${\bf \Gamma} (z)$ whose elements are given by \eqref{Gamma}. 
An important feature of this matrix is that the column vector ${\bf v}$ 
with elements $v_{kl}=\delta_{kl}$ is always left eigenvector of 
${\bf \Gamma} (z)$ with eigenvalue $z^{-1}$, i.e., 
$\sum_k \Gamma_{kk,k'l'} (z)=\delta_{k'l'}/z$ \cite{EPJB}. 
This equality ensures the conservation of the trace of the density 
matrix $\rho$, and follows simply from 
$\sum_k  \langle k | \mathrm{Tr}_{\cal E} (\ldots)   | k \rangle
=\mathrm{Tr} (\ldots)$. The matrix ${\bf \Gamma} (z)$ can thus be 
written as 
${\bf \Gamma} (z) = z^{-1} {\bf u}(z)  {\bf v}^T + {\bf \Gamma}' (z)$ 
where $ {\bf v}^T {\bf \Gamma}' (z)=0$ and $ {\bf v}^T {\bf u} (z)=1$. 
The column vector ${\bf u} (z)$ is right eigenvector of 
${\bf \Gamma} (z)$ with eigenvalue $z^{-1}$. Provided it has no pole 
on the real axis, the corresponding term of ${\bf \Gamma} (z)$ can be 
analytically continued in the lower half plane and gives a constant 
contribution to the time-evolved density matrix \eqref{rhot}. 
Since ${\bf v}^T {\bf r}=\sum_k r_{kk}=1$ for any density matrix 
$\rho(0)$,  this contribution does not depend on the initial state 
of ${\cal S}$. In summary, the steady state of the open system 
${\cal S}$ is $\sum_{k,l} u_{kl} | k \rangle \langle l |$ where 
$u_{kl}$ are the elements of the column vector ${\bf u}$ determined by 
\begin{equation}
\lim_{\eta \rightarrow 0^+} \left\{ i\eta {\bf \Gamma} (i\eta) \right\} 
{\bf u}= {\bf u}. 
\label{evpu}
\end{equation} 
Note that the condition $[\rho_{\cal E},H_{\cal E}]=0$ was not 
used to derive this equation.

\subsubsection{Weak coupling limit}

To determine the steady state of ${\cal S}$ in the limit of weak coupling 
to ${\cal E}$, we first expand the matrix elements \eqref{Gamma} 
in powers of the Liouvillian ${\cal L}_{int} \ldots =[H_{int},\ldots]$. 
We obtain 
\begin{equation}
\Gamma_{kl,k'l'} (z)=\frac{1}{z-\epsilon_{k}+\epsilon_{l} } \left\{ 
\delta_{k'k} \delta_{l'l} 
+  i \frac{\gamma_{kl,k'l'} (z)}{z-\epsilon_{k'}+\epsilon_{l'}}  \right\}
\end{equation}
up to second order, where $\gamma_{kl,k'l'} (z)$ 
can be expressed in terms of the correlation functions 
$C^{kl}_{k'l'} (t) = \mathrm{Tr}[\rho_{\cal E} \exp(itH_{\cal E}) h_{kl} 
\exp(-itH_{\cal E}) h_{k'l'}]$ of the environment operators 
$h_{kl}=\langle k | H_{int} | l \rangle=h_{lk}^\dag$, as
\begin{multline}
\gamma_{kl,k'l'} (z) = \int_0^\infty dt e^{izt} \Big\{ 
e^{it\omega_{l'k}} C^{l'l}_{kk'} (t) 
+ e^{it\omega_{lk'}} C^{l'l}_{kk'} (-t)  \Big. \\
\Big. - \sum_j \big[ \delta_{ll'} e^{it\omega_{l j}} C^{kj}_{jk'} (t) 
+ \delta_{kk'} e^{it\omega_{jk}} C^{l'j}_{jl} (-t) \big] \Big\} .
\label{petitgamma}
\end{multline} 
In this expression, we have used the notation 
$\omega_{kl}=\epsilon_k-\epsilon_l$. The stationarity of the correlation 
functions $C^{kl}_{k'l'}$ stems directly from the steady environment 
assumption $[\rho_{\cal E},H_{\cal E}]=0$. For the Hamiltonian 
\eqref{H} and with the definitions \eqref{def}, 
$h_{11}=-h_{44}=v_1+v_2$, $h_{22}=-h_{33}=v_1-v_2$, 
$h_{13}=h_{24}=w_1$, $h_{12}=h_{34}=w_2$ and 
$h_{14}=h_{23}=0$. 

In the absence of interaction between ${\cal S}$ and ${\cal E}$, 
the eigenvalue problem \eqref{evpu} reduces to 
$(\epsilon_k - \epsilon_l) u_{kl}=0$. Consequently, the only matrix 
elements $u_{kl}$ with nonvanishing zeroth-order approximations 
are that for which $\epsilon_k=\epsilon_l$. Thus, if the energy 
spectrum $\{\epsilon_k\}$ is nondegenerate, the corresponding 
steady density matrix is diagonal in the basis $\{| k \rangle \}$. 
In the opposite case, there can exist coherences between states 
$| k \rangle$ of equal energy. The matrix elements $u_{kl}$ to 
zeroth order, are determined by the equations
\begin{equation}
\sum_{\epsilon_{k'}=\epsilon_{l'}} \gamma_{kl,k'l'} (i0^+)  
u_{k'l'} = 0 
\label{gamma}
\end{equation}
where $k$ and $l$ satisfy $\epsilon_k=\epsilon_l$. The remaining 
coherences $u_{kl}$ are at least of first order in $H_{int}$. 
By writing explicitly the coefficients $\gamma_{kl,k'k'} (i0^+)$, 
it can be shown that, for an environment in thermal equilibrium, 
i.e., $\rho_{\cal E} \propto \exp(-H_{\cal E}/T)$, the thermal state 
$u_{kl}\propto \delta_{kl} \exp(-\epsilon_k/T)$ 
is solution of \eqref{gamma}, even in the presence of degeneracy 
in the spectrum of $H_{\cal S}$, see Appendix.

\subsection{Different splitting energies}\label{sec:Dse}

For unequal nonzero $\Delta_1$ and $\Delta_2$, 
the spectrum of the Hamiltonian 
$H_{\cal S}=-\sum_i \Delta_i \sigma_z^{(i)}/2$ is non degenerate. 
The zeroth-order steady state of the TLS is thus a statistical mixture 
of the states \eqref{def}. The relation \eqref{gamma} becomes
\begin{equation}
\left( \begin{array}{cccc} 
0 & -{\tilde \gamma}_1^- & 
-{\tilde \gamma}_2^-  & {\tilde \gamma}_1^+ +{\tilde \gamma}_2^+ \\
-{\tilde \gamma}_1^- & 0 & 
{\tilde \gamma}_1^+ +{\tilde \gamma}_2^- & -{\tilde \gamma}_2^+  \\ 
-{\tilde \gamma}_2^- & {\tilde \gamma}_1^- +{\tilde \gamma}_2^+ & 
0  & -{\tilde \gamma}_1^+  \\ 
{\tilde \gamma}_1^- +{\tilde \gamma}_2^- & 
-{\tilde \gamma}_2^+& -{\tilde \gamma}_1^+ & 0 
\end{array} \right) 
\left( \begin{array}{c} p_1 \\ p_2 \\ p_3 \\ p_4
\end{array} \right) = 0 \label{Diffspl}
\end{equation}
where $p_k=u_{kk}$. The elements of the above matrix can be 
written as
\begin{equation}
{\tilde \gamma}_i^{+/-} = 2\pi \sum_{A,B} P_{A/B} 
|\langle B | w_i | A \rangle |^2 \delta(E_A-E_B+\Delta_i) 
\label{gammatilde}
\end{equation}
where $E_A$ and $| A \rangle$ denote the eigenenergies and 
eigenstates of $H_{\cal E}$, and $P_A$ are the eigenvalues 
of $\rho_{\cal E}$. The coefficients ${\tilde \gamma}_i^+$ and 
${\tilde \gamma}_i^-$ are the Fermi golden rule rates of the 
TLS $i$ \cite{CDG}. 

The solution of \eqref{Diffspl} leads to a product steady state 
$\rho=\rho_{1}\otimes \rho_{2}$ where
\begin{equation}
\rho_i = ({\tilde \gamma}_i^+ +{\tilde \gamma}_i^-)^{-1} 
\big[ {\tilde \gamma}_i^+ |+\rangle_i {_i}\langle + | + 
{\tilde \gamma}_i^- |-\rangle_i {_i}\langle - | \big] . \label{rhon}
\end{equation}
The two TLS are uncorrelated, to lowest order in $H_{int}$, when 
their splitting energies are different from each other. Moreover, 
the steady state $\rho_i$ of TLS $i$ is the same in the presence 
or absence of the other TLS. In the special case $\Delta_2=0$, 
the zeroth-order coherences $u_{12}$ and $u_{34}$ are a priori 
different from zero since $\epsilon_1=\epsilon_2$ and 
$\epsilon_3=\epsilon_4$.  But, for an environment ${\cal E}$ 
consisting of several heat baths, it is shown in the Appendix 
that $\rho=\rho_{1} \otimes I/4$ where $\rho_1$ is given 
by \eqref{rhon} and $I$ is the $2\times 2$ identity matrix, 
is steady state. 

\subsection{Identical splitting energies}\label{sec:Ise}

For $\Delta_1=\Delta_2=\Delta \ne 0$, the states $| 2 \rangle$ 
and $| 3 \rangle$ have the same energy 
$\epsilon_2=\epsilon_3=0$. The other energies are 
$\epsilon_4=-\epsilon_1=\Delta$. Here, equation \eqref{gamma} 
takes the form
\begin{equation}
\left( \begin{array}{ccc} 
\mbox{\boldmath ${\tilde \gamma}$}  & 
-\mbox{\boldmath ${\beta}$}^*  
&-\mbox{\boldmath ${\beta}$}  \\
-\mbox{\boldmath ${\beta}$}^T & \alpha & 0 
\end{array} \right) 
\left( \begin{array}{c} {\bf p} \\ c \\ c^*
\end{array} \right) = 0 \label{Idespl}
\end{equation}
where $c=u_{23}=u_{32}^*$, \mbox{\boldmath ${\tilde \gamma}$} 
is the $4\times 4$ matrix given in \eqref{Diffspl} and 
${\bf p}^T=(p_1 \; p_2 \; p_3 \; p_4)$. The coefficient $\alpha$ 
reads as
\begin{multline}
\alpha =\frac{1}{2} \left( {\tilde \gamma}_1^- +{\tilde \gamma}_1^+ +
 {\tilde \gamma}_2^- +{\tilde \gamma}_2^+  \right)
 + \sum_{A,B} P_A |\langle A | v | B \rangle |^2 \delta(\omega_{AB}) 
  \label{alpha} \\
+ i \sum_{A,B} \big[ |\langle A | w_1 | B \rangle|^2 - 
|\langle A | w_2 | B \rangle|^2 \big] 
\frac{P_A+P_B}{\omega_{AB}-\Delta}
\end{multline}
where $v=2\sqrt{\pi} (v_1-v_2)$ and $\omega_{AB}=E_A-E_B$. 
The elements of $\mbox{\boldmath $\beta$}^T=
(\beta_1 \; \beta_2 \; \beta_3\; \beta_4)$ are given by
\begin{equation}
\beta_{1/4} = 2\pi \sum_{A,B} P_{A/B} \langle A | w_1 | B \rangle 
\langle B | w_2^{\dag} | A \rangle  \delta(\omega_{AB}-\Delta) , 
\label{beta14}
\end{equation}
$\beta_{2}=-(\beta_1+\beta_4)/2 + i{\tilde \beta}$ 
and $\beta_3=-(\beta_1+\beta_4)/2-i{\tilde \beta}$, where
${\tilde \beta} =  \sum_{A,B} \langle A | w_2^{\dag} | B \rangle 
\langle B | w_1 | A \rangle (P_A-P_B)/(\omega_{AB}+\Delta)$. 
For two identical two-level atoms coupled to the electomagnetic 
vacuum, $\beta_4$ and ${\tilde \beta}$ are, respectively, the collective 
decay rate and the dipole-dipole interaction energy of the atoms 
\cite{CKS,L}.  

It is instructive, for the following, to relate the coefficients 
\eqref{beta14} to Fermi golden rule rates. Instead of 
analysing the influence of ${\cal E}$ on the two TLS in 
the basis of product states \eqref{def}, the basis made up of 
the states $|1\rangle$, $|4\rangle$ and the entangled Bell states 
\begin{equation}
|\psi^\pm \rangle=\frac{| 2 \rangle \pm |3 \rangle}{\sqrt{2}}
=\frac{1}{\sqrt{2}}\big( | + \rangle_1 | - \rangle_2 \pm |
 - \rangle_1 | + \rangle_2 \big) , \label{Bs}
\end{equation}
can be used. Both bases correspond to the same energy 
spectrum $\{ \pm \Delta,0 \}$. The Fermi golden rule rates 
for the downward transitions $|4\rangle \rightarrow |\psi^{\pm} \rangle$ 
are given by $2\pi \sum_{A,B} P_{A} |\langle B | \langle \psi^{\pm} | 
H_{int} | 4 \rangle | A \rangle |^2 \delta(\omega_{BA}-\Delta)
=({\tilde \gamma}_1^{+}+{\tilde \gamma}_2^{+})/2
\pm \mathrm{Re} \beta_4$. This last expression is also valid 
for $|\psi^{\pm} \rangle \rightarrow |1\rangle$. For the upward 
transitions $|1\rangle \rightarrow |\psi^{\pm} \rangle$ and 
$|\psi^{\pm} \rangle \rightarrow |4\rangle$, the rates are 
$({\tilde \gamma}_1^{-}+{\tilde \gamma}_2^{-})/2 
\pm \mathrm{Re} \beta_1$.

Equation \eqref{Idespl} can be solved by diagonalizing 
\mbox{\boldmath ${\tilde \gamma}$}. 
The eigenvalues of  \mbox{\boldmath ${\tilde \gamma}$}  are 
$\lambda_0=0$, 
$\lambda_1={\tilde \gamma}_1^- +{\tilde \gamma}_1^+$, 
$\lambda_2={\tilde \gamma}_2^- +{\tilde \gamma}_2^+$ and 
$\lambda_4=\lambda_1+\lambda_2$. We denote by 
\mbox{\boldmath $\psi$}$_n$ and \mbox{\boldmath $\phi$}$_n$ 
the corresponding right and left eigenvectors. Since 
\mbox{\boldmath $\psi$}$_0$ is the only right eigenvector for 
which the sum of its elements does not vanish, 
${\bf p}=\mbox{\boldmath $\psi$}_0
+\sum_{n>0} \lambda_n^{-1} 
\mbox{\boldmath $\psi$}_n \mbox{\boldmath $\phi$}_n^T ( c 
\mbox{\boldmath $\beta$}^* + c^* \mbox{\boldmath $\beta$} )$, 
and the coherence $c$ is solution of 
\begin{equation}
\alpha c  \label{eqc}
-\mbox{\boldmath $\beta$}^T \mbox{\boldmath $\psi$}_0 
- \sum_{n>0} \lambda_n^{-1} \mbox{\boldmath $\beta$}^T
\mbox{\boldmath $\psi$}_n \mbox{\boldmath $\phi$}_n^T (  
\mbox{\boldmath $\beta$}^* c + \mbox{\boldmath $\beta$} c^* ) 
= 0 .
\end{equation}
We will see in the next section that $c$ can be nonzero 
and lead to stationary entanglement of the TLS. 
In the special case $\Delta_1=\Delta_2=0$, it can be shown 
that $c=0$ and $p_k=1/4$ are solutions of \eqref{Idespl}, 
if ${\cal E}$ is made up of heat baths, see Appendix. 

The treatment of section \ref{sec:Dse} applies when the difference 
$\delta=\Delta_1-\Delta_2$ is large enough that it can be considered 
finite in the expansion in terms of the interaction Hamiltonian 
$H_{int}$. In \eqref{Idespl}, this difference is exactly zero. A possible 
approach to understand the influence of a small $\delta$, consists in 
expanding the coefficients \eqref{Gamma} both in $H_{int}$ and 
$\delta$. This gives equation \eqref{Idespl} with $\alpha+i\delta$ 
in place of $\alpha$, which reduces to \eqref{Idespl} for $\delta$ 
much smaller than the other matrix elements, and leads to 
the uncorrelated state \eqref{rhon} with $\Delta_1=\Delta_2$, in 
the opposite limit. 

\section{Multiple heat baths environment}\label{sec:Bhb}

In this section, we consider an environment ${\cal E}$ made up of 
several heat baths, as sketched in Fig.\ref{fig:test}, each consisting 
of an infinite number of harmonic degrees of freedom which are 
coupled linearly to the TLS. In other words, 
the spin-boson model \cite{LCDFGZ}, which appropriately describes 
various physical environments \cite{QDS}, is generalized to two TLS 
and several heat reservoirs. We show that two bosonic baths can 
induce stationary entanglement of two identical TLS. 

\subsection{Environment model}\label{sec:Em}

We write the Hamiltonian of ${\cal E}$ as 
$H_{\cal E}=\sum_n H_{{\cal E}n}$ where $n$ runs over 
the heat baths and 
\begin{equation}
H_{{\cal E}n}=\sum_q \omega_{nq} a^\dag_{nq} a_{nq} .
\end{equation}
In this expression, the sum runs over the harmonic modes 
of the bath $n$. The annihilation operators $a_{nq}$ satisfy 
the bosonic commutation relation 
$[a_{nq},a^\dag_{n'q'}]=\delta_{nn'}\delta_{qq'}$. 
For the coupling operators, we consider 
\begin{equation}
w_i = \sum_{n,q} k_{nq}^{(i)} \left( a^\dag_{nq} + a_{nq} \right) , 
\label{boson}
\end{equation}
and a similar expression for $v_i$. The coupling parameters 
$k_{nq}^{(i)}$ are assumed to be real. The environment is initially 
in the state $\rho_{\cal E}\propto \prod_n \exp(-H_{{\cal E}n}/T_n)$ 
where $T_n$ is the temperature of bath $n$. 

Here, the rates \eqref{gammatilde} can be written as
\begin{equation}
{\tilde \gamma}_i^\pm=\pm \sum_n 
\frac{J_n^{(i)}}{1-e^{\mp \Delta_i/T_n}}
\label{tildegammahb}
\end{equation}
where $J_n^{(i)} = 2\pi \sum_{q} [ k^{(i)}_{n q} ]^2 
\delta(\omega_{n q}-\Delta_i)$, and the coefficients \eqref{beta14}, 
which are relevant only in the case $\Delta_i=\Delta$, are given by 
similar expressions with $J_n^{(i)}$ replaced by 
$K_n = 2\pi \sum_{q} k^{(1)}_{nq} k^{(2)}_{n q} 
\delta(\omega_{nq}-\Delta)$. Clearly, $J^{(i)}_n$ is necessarily 
positive but not $K_n$, and $|K_n|^2 < J^{(1)}_n J^{(2)}_n$. 
The main difference between $J^{(i)}_n$ and $K_n$ is that 
the former depends only on the coupling of TLS $i$ to ${\cal E}$, 
whereas the latter is determined by both coupling operators 
$w_1$ and $w_2$. An important physical parameter that controls 
the ratio $|K_n|^2/J^{(1)}_n J^{(2)}_n$ is the distance $d_n$ between 
the TLS coupling points to bath $n$. This ratio reaches its maximum 
value of $1$ when the two TLS interact in exactly the same way with 
bath $n$, which necessarily means $d_n=0$ \cite{ZR}. This spatial 
dependence is discussed more fully at the end of section \ref{sec:Ccf}.

Finally, we comment on the second term in \eqref{alpha}, which plays 
a role in the following. It can be cast into the form 
$\sum_n \int_0^{\infty} d\omega \delta(\omega) 
L_n(\omega)/\tanh(\omega/2T_n)$ where 
the spectral functions $L_n$ are defined similarly to $J^{(i)}_n$ 
with $\omega$ in place of $\Delta_i$. The function $L_n$ vanishes 
for frequencies $\omega$ higher than a cut-off frequency \cite{QDS}. 
Its low-frequency behavior leads to various possibilities. First, 
$\alpha$ is finite only if, for any $n$, $L_n/\omega$ does not diverge 
for $\omega \rightarrow 0$. If, in this limit, this ratio goes to zero 
for any $n$, then the second term in \eqref{alpha} vanishes. 
This term reads as $\alpha_n T_n + \alpha_{n'} T_{n'}+ \ldots$ 
for Ohmic spectral densities $L_m \sim \omega$, $m=n$, 
$n'$, \ldots \cite{QDS}. For a bath consisting of a $D$-dimensional 
continuous field, $L_n \sim \omega^D$ for $\omega \rightarrow 0$, 
and hence is Ohmic for $D=1$. However, note that, whereas 
$J_n^{(i)}$ and $K_n$ are determined by the transverse coupling 
operators $w_i$, the functions $L_n$ depend on the longitudinal 
coupling. Consequently, $L_n$ can in principle be made as small 
as we wish, irrespective of the transverse coupling strength.

\subsection{Steady state for identical two-level systems}

From now on, we consider the case of identical TLS splitting 
energies $\Delta_i=\Delta$, for which, as seen above, stationary 
TLS entanglement may exist. We further assume that the two TLS 
are coupled identically to the heat baths, i.e., 
$J_n^{(1)}=J_n^{(2)}=J_n$. This can hold for $w_1 \ne w_2$ if 
the two TLS are connected to different points of bath $n$. 
As a consequence of these assumptions, 
${\tilde \gamma}_2^\pm={\tilde \gamma}_1^\pm$, 
see \eqref{tildegammahb}. To simplify the following expressions, 
we introduce the notations :
\begin{eqnarray}
 {\tilde \gamma}_1^+={\tilde \gamma}_2^+ = \gamma &,& 
{\tilde \gamma}_1^-={\tilde \gamma}_2^- = \gamma \eta  
\label{notations} \\
 \beta_4 = \beta &,&\beta_1 =  \beta \eta'   \;\; , \;\;  
\alpha = \gamma (1+\eta+ \xi) \nonumber
\end{eqnarray} 
where $\beta$, $\eta'$ and $\xi$ are real for the coupling 
operators \eqref{boson} and with the above assumptions. 
As discussed above, $\xi$ is determined by the longitudinal 
coupling, whereas all the other parameters are related to 
the lateral coupling operators $w_i$. The coefficient 
${\tilde \beta}$, defined right after \eqref{beta14}, is also 
real and does not contribute to the TLS steady state.

Under the assumption of real $\beta$, $\eta'$, $\xi$ 
and ${\tilde \beta}$, we find, from \eqref{eqc}, a real coherence
\begin{multline}
c = \frac{\gamma}{\beta}(\eta'-\eta)(1-\eta^2)\Big\{ 4(\eta'-\eta)^2
+ \frac{\gamma^2}{\beta^2} (1+\eta)^3(1+\eta+\xi) \Big. \\
\Big. -(1+\eta')(1+3\eta^2+3\eta'+\eta'\eta^2) \Big\}^{-1} . \label{csc}
\end{multline} 
The populations of the TLS steady state $\rho$ can be written in 
terms of $c$ as
\begin{multline}
\left( \begin{array}{c} p_1 \\ p_2 = p_3 \\ p_4
\end{array} \right) = \frac{1}{z^2} \left( \begin{array}{c} 1 \\ \eta \\ \eta^2
\end{array} \right)  +\frac{c\beta}{\gamma z^3} \Bigg\{  z(\eta+\eta')  
\left( \begin{array}{c} 1  \\ -1 \\ 1 \end{array} \right) \Bigg. \\
\Bigg.  +  (\eta'\eta+\eta'-\eta-1) \left( \begin{array}{c} -2  \\ 1-\eta \\ 
2\eta\end{array} \right) \Bigg\} \label{psc}
\end{multline} 
where $z=1+\eta$. Note that $c=0$ and $\rho$ is uncorrelated 
for $\eta'=\eta$ or $\eta=1$. When this last equality is satisfied, 
$\rho$ is proportional to the identity matrix, 
as expected from the case $\Delta=0$, see Appendix. The equality 
$\eta'=\eta$ holds, for instance, when ${\cal E}$ is in thermal 
equilibrium. The denominator in \eqref{csc} vanishes for 
$\gamma=\beta$, $\eta'=\eta$ and $\xi=0$. These three conditions 
are fulfilled for $w_2=w_1$ and $v_2=v_1$. There is not a unique 
steady state when the two TLS interact with ${\cal E}$ in exactly 
the same way \cite{ZR}. 
We also remark that, since $c$ is real and $p_2=p_3$, the 
TLS steady state can be written as 
$\rho=p_1 | 1 \rangle \langle 1 | + p_4 | 4 \rangle \langle 4 |+
(p_2+c) | \psi^+ \rangle \langle \psi^+ | 
+ (p_2-c) | \psi^- \rangle \langle \psi^- |$ 
with the Bell states $|\psi^\pm \rangle$ given by \eqref{Bs}.  
We wil see below that, though $|\psi^+ \rangle$ and $|\psi^- \rangle$ 
have the same energy $\epsilon_2=\epsilon_3=0$, there exists 
a parameter regime in which $p_4=0$ and $p_2=|c|$, and $\rho$ is 
hence entangled.

\subsection{Entanglement induced by two heat baths} 

We now study the entanglement of $\rho$ for an environment 
${\cal E}$ that consists of two heat baths of temperatures $T_1$ 
and $T_2$. The steady state $\rho$ is entangled if and only if 
its partial transpose 
$ \rho^\Gamma = \sum_k p_k | k \rangle \langle k |
 + c ( | 1 \rangle \langle 4 | + | 4 \rangle \langle 1 | )$ 
has negative eigenvalues \cite{P,HHH}. The eigenvalues of 
 $\rho^\Gamma$ are $p_2=p_3$ and
$\lambda_{\pm} = (p_1+p_4)/2 \pm [(p_1-p_4)^2 + 4 c^2]^{1/2}/2$. 
Clearly, only $\lambda_-$ can be negative. 

\subsubsection{Low-temperature entanglement region}

\begin{figure}
\centering \includegraphics[width=0.45\textwidth]{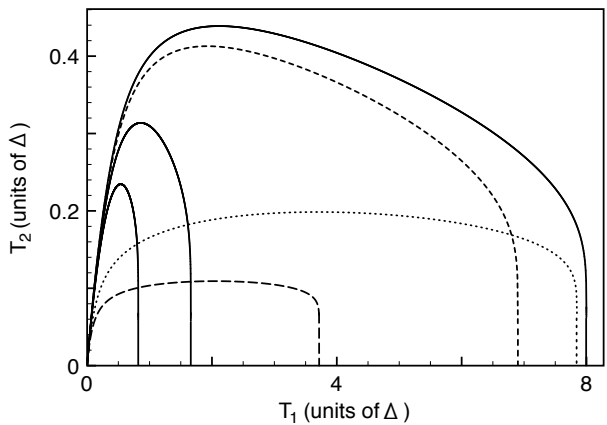}
\caption{\label{fig:essr0} Entanglement region in the $(T_1,T_2)$ 
plane for $K_1=0$. The TLS steady state is entangled for 
temperatures below the drawn line. The solid lines correspond to 
$K_2=J_2$, $\xi=0$, and $J_2/J_1=5$, $10$ and $50$. The size 
of the entanglement region increases with $J_2$. For the dashed 
and dotted lines, the coupling parameters are $\xi=0$, and, 
respectively, $K_2=0.72 J_2$ and $J_2=500 J_1$, and 
$K_2=0.8 J_2$ and $J_2=150 J_1$. The short-dashed line 
is obtained for an Ohmic $\xi=0.01 T_1/\Delta$ and 
for $K_2=J_2=50J_1$. }
\end{figure}

As an interesting example, we consider the case $\xi=0$ and 
$K_1=0$. This last condition means that the indirect interaction 
between the TLS is mediated only by bath $2$. With this value 
of $K_1$, the results discussed here hold also for the three 
bath setup depicted in Fig.\ref{fig:test} when $T_3=T_1$. 
We find that there can be a low-temperature region, determined 
by $J_2/J_1$ and $K_2/J_2$, in which $\rho$ is entangled, 
see Fig.\ref{fig:essr0}. We remark that the line delimiting 
this entanglement region in the $(T_1,T_2)$ plane, is tangent 
to the equilibrium line $T_1=T_2$ for $T_1,T_2 \rightarrow 0$, 
and is essentially vertical at its other end for $T_2 \ll \Delta$. 
These two behaviors come from the fact that the temperatures 
contribute to $\rho$ only via Boltzmann factors 
$\exp(-\Delta/T_n)$. 

Analytical results can be obtained by expanding the eigenvalue 
$\lambda_-$ to lowest order in these factors. It assumes negative 
values in the vicinity of $T_1=T_2=0$, for 
$|K_2| > J_2/\sqrt{2}$ and $J_2 > J_1 [\sqrt{2}|K_2|/J_2-1]^{-1}$. 
These requirements are the same in the Ohmic case 
discussed at end of \ref{sec:Em}, for which 
$\xi= {\bar \xi}_1 T_1 + {\bar \xi}_2 T_2$ 
vanishes in the limits $T_1,T_2 \rightarrow 0$. 
For given coupling parameters satisfying the above conditions, 
$\rho$ is not entangled if the temperatures $T_1$ and $T_2$ 
are too high. However, for $\xi=0$, the maximum possible value of 
$T_1$ is proportional to $\Delta J_2/J_1$ in the large $J_2$ limit, 
see Fig.\ref{fig:essr0}. Consequently, in this case, entangled states 
exist for any temperature $T_1$. For $T_2$, in contrast, 
our numerical results suggest that the steady state is always 
separable for $T_2$ greater than a value of about $0.567 \Delta$. 
Entangled states can be observed close to this temperature 
in the limit of large $J_2$. For 
$\xi= {\bar \xi}_1 T_1+ {\bar \xi}_2 T_2$, $\rho$ is necessarily 
separable for $T_1$ higher than a temperature that diverges for 
${\bar \xi}_1 \rightarrow 0$. 
  
\subsubsection{Requirements on the characteristics of 
the environment}\label{sec:Ccf}  

\begin{figure}
\centering \includegraphics[width=0.45\textwidth]{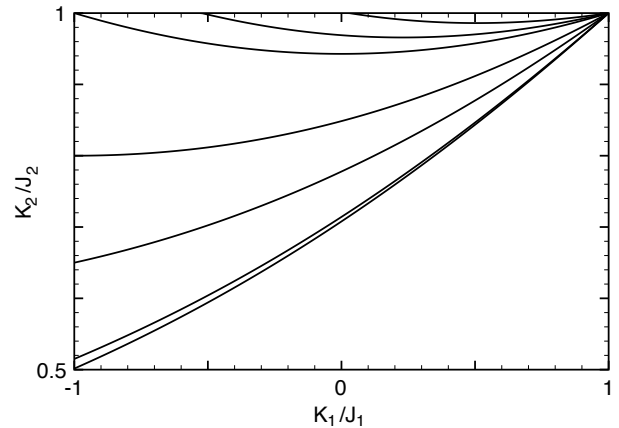}
\caption{\label{fig:essKJ} Region of the parameter plane 
$(K_1/J_1,K_2/J_2)$, $K_2>0$, where entangled steady states 
can be found, for $J_2/J_1=2.4$, $2.7$, $3$, $5$, $10$, $100$ and 
$1000$. The entanglement region is above the drawn line. Its size 
increases with $J_2$, to a maximum asymptotic value which is 
practically reached for $J_2=1000J_1$.}
\end{figure}
  
Stationary entanglement can also be obtained for $K_1 \ne 0$. 
Since $\rho$ is obviously invariant under the bath 
permutation $(J_1,K_1) \leftrightarrow (J_2,K_2)$, it is enough to 
consider $J_2>J_1$. In this case, it can be shown that there exist 
entangled steady states in the vicinity of $T_1=T_2=0$ if
\begin{equation}
\left( J_1+J_2 \right)^2-\left( K_1 + K_2 \right)^2 
- \left| K_1+K_2 \right| \left| K_2- K_1 \frac{J_2}{J_1} \right| < 0 .
\end{equation}
This condition remains the same if the signs of both $K_1$ and 
$K_2$ are changed. Figure \ref{fig:essKJ} shows, for $K_2>0$, 
the coupling parameter region where $\rho$ can be found 
entangled. The following interesting conclusions can be drawn 
from these results. There is a particular value of $J_2/J_1$ 
below which $\rho$ is separable. In other words, the couplings 
to the two heat baths must differ enough from each other in 
order to observe stationary entanglement. For given $J_2/J_1$ 
and $K_1/J_1$ such that entangled steady states exist, 
these states are obtained for $|K_2|/J_2$ not too far from $1$.

As mentioned above, the ratio $|K_2|/J_2$ depends essentially on 
the distance $d_2$ between the two points of bath $2$ where 
the TLS are connected. More precisely, it is determined by 
a dimensionless parameter ${\bar d}=\Delta d_2/v$ where $v$ 
is a characteristic field velocity of bath $2$. The ratio $|K_2|/J_2$ 
is small for large ${\bar d}$. This imposes limitations on $\Delta$ 
and on the temperature $T_2$ to obtain an entangled steady state. 
A distance $d_2$ of $1 \mu$m and a low field velocity $v$ of 
$10^3$ m.s$^{-1}$, which is the order of magnitude of the sound 
velocity in solids, give a temperature of about $10$ mK, which is 
an experimentally accessible value. Another important characteristic 
of bath $2$ is its dimensionality $D$. For example, for a continuous 
free field, $K_2/J_2$ is equal to $\sin ({\bar d})/{\bar d}$ for $D=3$, 
$J_0({\bar d})$ where $J_0$ is the zeroth order Bessel function 
of the first kind, for $D=2$, and $\cos ({\bar d})$ for $D=1$. Thus, 
in this last case, stationary entanglement can be obtained for large 
distances $d_2$ and the limitations discussed above do not apply.

\subsubsection{Maximum attainable entanglement}
  
\begin{figure}
\centering \includegraphics[width=0.45\textwidth]{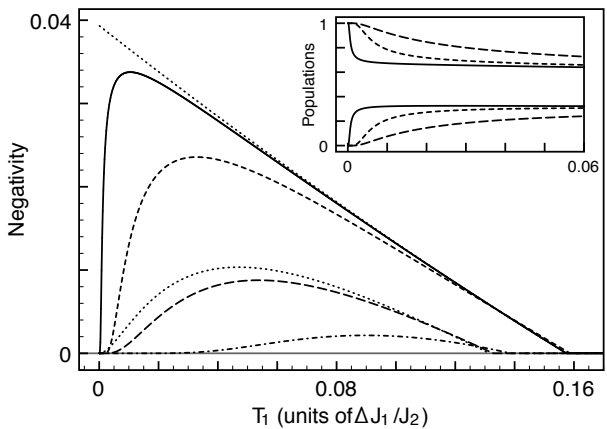}
\caption{\label{fig:Neg} Negativity as a function of $T_1$ in units 
of $\Delta J_1/J_2$ for $K_2=J_2=100J_1$ (short-dashed line), 
$K_2=J_2=1000J_1$ (full line), $K_2=0.95J_2=9.5J_1$ 
(dash-dotted line) and $K_2=0.95J_2=95J_1$ (dashed line). 
The dotted lines correspond to the large $J_2$ approximation 
discussed in the text. The inset shows the populations of 
the ground and singlet states 
as functions of $T_1$ in units of $\Delta J_1/J_2$ 
for $K_2=0.95J_2=95J_1$, $K_2=J_2=100J_1$ and 
$K_2=J_2=1000J_1$. The two other populations are small. 
The other parameters are $\xi=0$, $K_1=0$ and 
$T_2=0.1\Delta$.}
\end{figure}

Finally, we present quantitative results for the entanglement of 
the steady state $\rho$. As a measure of entanglement, we use 
the negativity ${\cal N}(\rho)= (\Vert \rho^\Gamma \Vert_1-1)/2$ 
where $\Vert . \Vert_1$ denotes the trace norm \cite{ZHSL,VW}. 
Negativity ranges from $0$ for separable states to $1/2$ for 
maximally entangled states. Here, it is equal to $-\lambda_-$ 
when this eigenvalue is negative, and to $0$ otherwise. 
The maximum value of ${\cal N}$ that we have found, 
is reached for the coupling parameters $K_1=0$, $\xi=0$ 
and $J_2 \gg J_1$, and the temperatures $T_2 \ll \Delta$ and 
$T_1 \gg \Delta$, see Fig.\ref{fig:Neg}. In this regime, the TLS 
steady state is given by
\begin{eqnarray}
p_1=\sigma \left[ (1+\theta)^2(1+2\theta) -\kappa^2 \right] &,& 
p_2=\sigma \theta (1+\theta)(1+2\theta) \nonumber \\
p_4=\sigma \theta^2 (1+2\theta) &,& c = -\sigma \kappa \theta
\end{eqnarray}
where $\sigma=[(1+2\theta)^3 -\kappa^2]^{-1}$, $\kappa=K_2/J_2$ 
and $\theta=(T_1/\Delta)(J_1/J_2)$. For $|\kappa| \ne 1$,
as $\theta$ increases from zero to infinity, the ground state population 
$p_1$ decreases from $1$ to $1/4$, $p_2=p_3$ and $p_4$ increases 
from $0$ to $1/4$, and $|c|$ increases from zero to a maximum and 
then decays back to zero. The low $\theta$ behavior is very different 
for $|\kappa|=1$. In this case, $c$ is finite in the limit 
$\theta \rightarrow 0$. For $K_2=\pm J_2$ and a temperature 
$\Delta \ll T_1 \ll \Delta J_2/J_1$, 
we find $\rho = (2/3) | + \rangle_1 {_1\langle + |} \otimes 
 | + \rangle_2 {_2\langle + |}
 +(1/3) | \psi^{\mp} \rangle \langle \psi^{\mp} |$
with the Bell states $| \psi^{\mp} \rangle$ given by \eqref{Bs}, 
and a negativity ${\cal N} = (\sqrt{5}-2)/6 \simeq 0.04$. 
The same entangled state can be reached for $K_1 \ne 0$, as it will 
be clear from the discussion below. Our numerical results suggest that 
finite values of ${\cal N}$ correspond generally to states $\rho$ such 
that essentially only the ground state and one of the Bell states 
\eqref{Bs} are populated, see inset of Fig.\ref{fig:Neg}. 

To better understand the above results, it is interesting to consider 
the rates discussed after \eqref{Bs}. For $T_2 \ll \Delta$ and 
$T_1 \gg \Delta$, the rates of the upward transitions 
$|1\rangle \rightarrow |\psi^{\pm} \rangle$ and 
$|\psi^{\pm} \rangle \rightarrow |4\rangle$ are 
$r_{up}^{\pm}=(J_1\pm K_1) T_1/\Delta$, and that of 
the downward transitions 
$|4\rangle \rightarrow |\psi^{\pm} \rangle$ 
and $|\psi^{\pm} \rangle \rightarrow |1\rangle$ are 
$J_2 \pm K_2+r_{up}^{\pm}$. For 
$K_2=J_2$ and $T_1 \ll \Delta J_2/J_1$, the rate 
of $|4\rangle \rightarrow |\psi^- \rangle$ and 
$|\psi^- \rangle \rightarrow |1\rangle$ is equal to 
$r_{up}^-$, whereas that of 
$|4\rangle \rightarrow |\psi^+ \rangle$ and 
$|\psi^+ \rangle \rightarrow |1\rangle$ is 
$2J_2 \gg r_{up}^{\pm}$.  Consequently, the states $|4\rangle$ 
and $|\psi^+ \rangle$ are essentially not populated, 
and the transition rate from the state $|\psi^- \rangle$ to 
the ground state $|1\rangle$ is effectively twice that of 
the reverse transition, leading to a factor of two between 
the two corresponding populations. The situation is similar 
fo $K_2=-J_2$. If $|K_2|$ is too 
far from $J_2$ or if $T_1$ is too high, the values of the different 
rates are comparable and so are the populations of the states 
$|1\rangle$, $|\psi^{\pm} \rangle$ and $|4\rangle$, and hence 
$\rho$ is separable. 

\section{Conclusion}

In summary, we have studied a system of two independent 
TLS weakly coupled to a stationary non-equilibrium environment. 
Considering first a general open system, we have determined 
their steady state. Without specifiying any further 
the surroundings of the TLS, it can be shown that their steady state 
is uncorrelated if their splitting energies are different from each 
other. Moreover, the state of each TLS is the same wether or not 
the other TLS is present. Consequently, in this case, a finite 
strength of the coupling to the environment is required 
to possibly generate stationary TLS entanglement. 
In the opposite case of identical splitting energies, on the contrary, 
stationary correlations between the TLS can exist for extremely weak 
coupling to the environment.

To determine wether these correlations can be quantum, we have 
considered the case of an environment consisting of several bosonic 
heat baths at different temperatures. We have shown that, for TLS 
coupled similarly to two baths, there are temperatures and coupling 
parameters for which the TLS steady state is entangled. An important 
requirement is that, for at least one bath, the points to which the TLS 
are connected must be close enough to each other. However, 
this condition can be relaxed when one of the bath is one-dimensional. 
In this case, the TLS can be as far apart as we like. There are also 
requirements on the baths temperatures. Essentially, one of them 
must be sufficiently low, of the order of the TLS splitting energy. 
Depending on the characteristics of the coupling, the other 
temperature can be unlimited. 

We have found a parameter regime where the TLS steady state is 
a statistical mixture of the product ground state and of the entangled 
singlet state with weights 2/3 and 1/3, respectively. This mixed state 
is entangled and the corresponding negativity is about 0.04 which is 
the largest value we have obtained. Interestingly, this regime can be 
fully understood in terms of Fermi golden rule transitions between 
appropriate states. To conclude, our results show that a relatively 
simple non-equilibrium environment can lead to stationary 
entanglement of two TLS, but certainly do not exhaust all 
the possible effects of stationary non-equilibrium surroundings 
on quantum correlations. Larger entanglement of independent TLS, 
as measured by negativity for instance, may be achievable with 
other environments or for TLS coupled differently to the environment. 
Further studies in these directions would be of interest. 

\begin{appendix}

\section{Special uncorrelated states}

In this appendix, our purpose is to show that, for some special cases, 
the solution of \eqref{gamma} is of the form 
$u_{kl}=p_k \delta_{kl}$. This is the case if the sums
\begin{multline}
\sum_{k'} \gamma_{kl,k'k'} (i0^+) p_{k'} =  \pi \sum_{A,B,k'} P_A 
\langle A | h_{kk'} | B \rangle \langle B | h_{k'l} | A \rangle \\
\times \left[ \left(p_k+p_l-2\frac{P_B}{P_A}p_{k'} \right) 
\delta(\omega_{AB}-\omega_{k'k}) 
+\frac{i}{\pi} \frac{p_l-p_k}{\omega_{AB}-\omega_{k'k}} \right] \\  
\label{sum}
\end{multline}
vanish for $k$ and $l$ such that $\epsilon_k=\epsilon_l$.

For an environment in thermal equilibrium, i.e., 
$P_A \propto \exp(-E_A/T)$ where $T$ is its temperature, 
$u_{kl} \propto \exp(-\epsilon_k/T) \delta_{kl}$ satisfies 
\eqref{gamma} since, in the sums \eqref{sum}, $p_k=p_l$ and 
$P_B p_{k'}/ P_A p_k = \exp[-(\omega_{BA}+\omega_{k'k})/T]$. 
This proof applies to any system ${\cal S}$.

We now consider the case of zero splitting energy $\Delta_2$ 
and of an environment ${\cal E}$ that consists of heat baths at 
different temperatures $T_n$. First, the populations $p_k$ obtained 
in section \ref{sec:Dse} ensure the vanishing of \eqref{sum} for $k=l$. 
For $k \ne l$, we start by showing that 
${\tilde \gamma}_2^+={\tilde \gamma}_2^-$ which implies $p_1=p_2$ 
and $p_3=p_4$, see \eqref{rhon}. The difference of these rates 
reads as
\begin{equation}
{\tilde \gamma}_2^+ - {\tilde \gamma}_2^- = 
2\pi \sum_{A,B} (P_A-P_B) |\langle B | w_2 | A \rangle |^2 
\delta(\omega_{AB}) . 
\label{diff}
\end{equation}
For the kind of environment considered, 
$H_{\cal E}=\sum_n H_{{\cal E}n}$ 
and hence its eigenstates and eigenenergies can be written as 
$| A \rangle = \prod_n | A^{(n)} \rangle$ and 
$E_A = \sum_n E_{A^{(n)}}$. The populations $P_A$ factorise as 
$P_A \propto \prod_n \exp(-E_{A^{(n)}}/T_n)$. The TLS are 
coupled to each bath thus $w_2=\sum_n w_{2n}$. Consequently, 
the difference \eqref{diff} satisfies
\begin{multline}
{\tilde \gamma}_2^+ - {\tilde \gamma}_2^- \propto 
\sum_n \sum_{A^{(n)},B^{(n)}} |\langle B^{(n)} | w_{2n} | 
A^{(n)} \rangle |^2 \\ 
\times \left(e^{-E_{A^{(n)}}/T_n}-e^{-E_{B^{(n)}}/T_n}\right)  
\delta(E_{A^{(n)}}-E_{B^{(n)}}) 
\end{multline}
and hence vanishes. For $\Delta_2=0$, the sum \eqref{sum} 
must be zero for $(k,l)=(1,2)$, $(2,1)$, $(3,4)$ and $(4,3)$. 
For these cases, the equalities $p_1=p_2$ and $p_3=p_4$, 
shown above, lead to 
\begin{multline}
\sum_{k'} \gamma_{kl,k'k'} (i0^+) p_{k'} =  2\pi \sum_{A,B,k'}
\langle A | h_{kk'} | B \rangle \langle B | h_{k'l} | A \rangle \\
\times \left(p_k P_A -P_B p_{k'} \right) 
\delta(\omega_{AB}-\omega_{k'k})  .
\end{multline}
For the Hamiltonian \eqref{H}, 
$h_{14}=h_{23}=0$ and hence the only terms that contribute 
to the above sum are such that $p_{k'}=p_k$ and 
$\omega_{kk'}=0$. Thus, it vanishes for the same reasons 
as \eqref{diff} does. 

For $\Delta_1=\Delta_2=0$, the sum \eqref{sum} must be zero 
for any $(k,l)$. In this case, for an environment that consists 
of heat baths, $u_{kl}=\delta_{kl}/4$ satisfies 
\eqref{gamma} since $\sum_{k'} \gamma_{kl,k'k'} (i0^+) = 0$. 

\end{appendix}

\end{document}